\documentclass[final,3p,times,twocolumn]{elsarticle}

\usepackage{amsmath}
\usepackage{amssymb}
\usepackage{graphicx}
\usepackage{bm}
\usepackage{color}

\begin{document}

\begin{frontmatter}

\title{Proposal for locality test of the Aharonov-Bohm effect 
 via Andreev interferometer without a loop
 }

\author{Kicheon Kang}
\ead{kicheon.kang@gmail.com}
\address{Department of Physics, Chonnam National University,
 Gwangju 500-757, Republic of Korea}
%

\begin{abstract}
We propose a quantitative test of the quantum locality in 
the electromagnetic interaction that generates the Aharonov-Bohm effect.
For this purpose, we analyze the Lorentz-covariant field interaction approach
based on the local action of gauge-invariant quantities only (``local" theory), 
and compare it with
the standard potential-based (``nonlocal") theory. 
Whereas the two approaches yield identical results for topological phase and
any phenomenon involving classical equation of motion,
an example violating this equivalence is presented; 
interference of the Andreev reflections from two independent superconducting
inputs into a single normal metallic output. 
A well-defined phase shift of the interference is predicted 
in the ``local" theory. In contrast, 
the potential-based Lagrangian fails the corresponding prediction.
This result is significant
as it can settle the issue of quantum locality in
the electromagnetic interaction.
\end{abstract}
 

\end{frontmatter}

%
%

\section{Introduction}
%
Nonlocality is one of the most characteristic features of quantum theory. 
The Einstein-Podolsky-Rosen (EPR) paradox~\cite{einstein35} and violation of 
the Bell's 
inequality~\cite{bell64} demonstrate that nonlocal correlations 
occur in quantum entanglement
that cannot be accounted for by any local realistic theory. 
This ``EPR nonlocality" has been the subject of intensive study for several
decades (see e.g., Ref.~\cite{greenstein06}). In addition, understanding
and utilizing this nonlocality is one of the greatest 
achievements in quantum information science.
The EPR nonlocality is purely kinematic in that no dynamics governed by the
quantum equation of motion is involved.
Another kind of nonlocality, which manifests in the form of the 
Aharonov-Bohm (AB) effect~\cite{aharonov59}, also appears in quantum theory.
This type of nonlocality is dynamic; it is
the nonlocality of a quantum equation of motion~\cite{popescu10}, and
has received less attention. 
Even though the AB interference is well described by the standard quantum
theory and has been experimentally confirmed, 
a counterintuitive nonlocality is required;
this prevents a causal explanation of this phenomenon. 
The AB effect is understood in terms of the electromagnetic potentials 
(which are gauge dependent and, therefore, cannot be physical fields) and/or 
the ``nonlocality" of the interaction between a moving charge and 
electromagnetic fields. In any case, the term ``AB nonlocality" 
indicates that the AB effect cannot be described
by local and causal action of gauge-invariant quantities.  
Unlike the case of EPR nonlocality, 
AB nonlocality has not been intensively investigated.
This primarily is because of the
the absence of a quantitative criterion for the experimental test 
(see also Y.~Aharonov~{\em et al.}~\cite{aharonov15}).

In this Letter, we propose an unambiguous test
of the locality in the AB effect. 
First, we point out that a general local realistic theory can be constructed
based on the Lorentz-covariant field interaction~(LCFI)~\cite{kang13,kang15}. 
The two approaches, the LCFI and the potential-based ones, 
yield the same results
for classical equations of motion and
for quantum phenomena involving topological phases.
Second, more importantly, we propose an intriguing example 
that breaks the equivalence of the two theories, i.e.,
an interferometer composed of two sources 
that does not form a closed loop in the particle paths. 
This is an analogue of the optical 
Pfleegor-Mandel interferometer~\cite{pfleegor67} (see also Fig.~1(a) of
Ref.~\cite{mandel99})
applied to charged particles with an external magnetic flux~(Fig.~1).
We show that the two theories lead to
different predictions for the interference fringe in this setup. 
The nonlocal Lagrangian does not yield a well-defined phase
shift induced by the localized flux, as there is no closed loop between
the two paths. On the other hand, with the local approach, 
the relative phase shift of the two
paths is obtained as a result of the accumulation of the  
local interaction.
Therefore, an observation of this type of interference 
excludes the standard approach with gauge-dependent potential, 
and thereby resolves the question of dynamical quantum locality in the 
electromagnetic interaction. 

%
\section{Nonlocal vs. local Lagrangians}
Our criterion for the locality is
that the Lagrangian is given by sum of the local overlap 
of gauge-independent quantities only.
This ``dynamical locality" provides a local realistic description
of quantum equations of motion, 
which is not satisfied in the potential-based theory~\cite{popescu10}.
Normally, a moving charge $q$ 
under the external electromagnetic field is 
described by the vector and the scalar potentials,
$\mathbf{A}(\mathbf{r},t)$ and $V(\mathbf{r},t)$, respectively,
at position $\mathbf{r}$ and time $t$. The Lagrangian of the system is given
by
\begin{subequations}
\label{eq:L}
\begin{equation}
 L = L_0 + L_{int} \,,  
\end{equation}
where $L_0=m\dot{\mathbf{r}}^2/2$ ($m$ is the mass) is the kinetic part, 
and the interaction 
is given by the Lorentz-covariant form~\cite{jackson62}, 
\begin{equation}   
 L_{int} = \frac{q}{c} \dot{\mathbf{r}}\cdot \mathbf{A} - qV \,.
\label{eq:Lint0}
\end{equation}
\end{subequations}
In this framework, the charge $q$ interacts locally with the
potentials, $\mathbf{A}$ and
$V$, but these quantities are gauge dependent, and thus does not fulfill
the locality criterion mentioned above. It is widely believed that this
is a unique approach for describing the quantum electromagnetic interaction
and that the potentials are indispensable~\cite{peshkin89}. 

In contrast to this common notion, a fully local theory was developed
based on the Lorentz-covariant field interaction~(LCFI)~\cite{kang13,kang15}.
The essence of the LCFI approach is that the influence of 
the external electric ($\mathbf{E}$) and magnetic ($\mathbf{B}$)
fields on a moving charge $q$ is represented, without potentials, 
by the Lorentz-covariant Lagrangian 
\begin{subequations}
 \label{eq:Lintf} 
\begin{equation}
 L_{int}^f = \frac{1}{4\pi} \int (\mathbf{B}_q\cdot\mathbf{B}
           - \mathbf{E}_q\cdot\mathbf{E}) \,d^3\mathbf{r}' \,,
\end{equation}
where $\mathbf{E}_q$($\mathbf{B}_q$) is the electric(magnetic) field produced
by the moving charge.
By adopting the relation 
$\mathbf{B}_q = \frac{1}{c}\dot{\mathbf{r}}\times\mathbf{E}_q$,
we obtain an instructive form,
\begin{equation}
 L_{int}^f = \dot{\mathbf{r}}\cdot \mathbf{\Pi}_q - U_q, 
\label{eq:Lint2}
\end{equation}
where
\begin{equation}
 \mathbf{\Pi}_q = \frac{1}{4\pi c} \int\mathbf{E}_q\times\mathbf{B} \, 
                d^3\mathbf{r}'
\label{eq:Pi}
\end{equation}
is the field momentum produced by the overlap between 
$\mathbf{E}_q$
and $\mathbf{B}$,
and
\begin{equation}
 U_q = \frac{1}{4\pi} \int \mathbf{E}_q\cdot\mathbf{E}\,d^3\mathbf{r}'
\label{eq:U}
\end{equation}
\end{subequations}
is the interaction energy stored in the electric fields.

Notably, 
the interaction Lagrangian $L_{int}^f$ of Eq.~\eqref{eq:Lintf} reproduces
the well-known topological phases derived from the conventional one, $L_{int}$
of Eq.~\eqref{eq:Lint0}. 
For instance, in a typical AB interferometer under an external magnetic field, 
the magnetic AB phase shift is given by an integral over a closed path
\begin{equation}
 \phi_{AB} = \frac{1}{\hbar} \oint \mathbf{\Pi}_q \cdot d\mathbf{r} 
           = \frac{q}{\hbar c} \oint \mathbf{A} \cdot d\mathbf{r} 
           = \frac{q\Phi}{\hbar c} \,,
\label{eq:phi-AB}
\end{equation}
where $\Phi$ is the magnetic flux threaded by the loop.
In addition, the classical equation of motion can be derived from
the ``local" Lagrangian, $L_f = L_0 + L_{int}^f$, which is also equivalent
to that obtained from the potential-based Lagrangian
(see Ref.~\cite{kang13} for details).

\section{Interferometric locality test}
Given the equivalence of the predicted results from the two different
approaches, the essential question is 
whether one can find any observable phenomenon that can discriminate the
LCFI theory from the potential-based one. As discussed above, the
two theories predict the same phase shift as far as interference with
a closed loop is concerned. Most of the quantum interferometers realized 
up to the present
belong to this class, and it may seem unlikely that the
equivalence is broken in reality
%
%

Here, we suggest a counter example, which adopts a different type
of interference that does break the equivalence.
Before we introduce a realistic setup involving 
superconducting hybrid junctions later,
let us first consider a prototype interferometer of a 
single charge ``coproduced" from two independent, 
but synchronized, sources
with a localized magnetic flux and a detector screen~(Fig.~1). 
At this point, our discussion is only schematic containing
the essential physics.
The main feature of this setup is that 
an interference is predicted 
as a function of the relative phase produced by the two paths 
without forming a closed loop. 
In fact, this is an analogue of the optical Pfleegor-Mandel 
interferometer~\cite{pfleegor67,mandel99}. 
The ``Pfleegor-Mandel" interference of single 
photons from two independent sources 
was recently shown very clearly at microwave frequencies 
(see Fig.~3(d) of Ref.~\cite{lang13}). 
The origin of that interference can be attributed
to the indistinguishability of the particle paths: 
one cannot tell which source has produced the
particle detected at a given point in the screen. 
A new aspect in our case is the introduction of charged particles, 
and thus the interference is also affected by the interaction
with the external magnetic field.

In the interferometer illustrated in Fig.~1, each source, $S_j$ ($j=1,2$),
simultaneously emits a charge state of $(u_j+v_jc_j^\dagger)|0\rangle$, 
involving superposition of the vacuum ($|0\rangle$) and a single charge. 
(Note that a normal electron cannot be in this state.)
The operator $c_j^\dagger$($c_j$) creates (annihilates) a particle at $S_j$.
The state of the entire system upon injection is given by
\begin{equation}
 |\psi\rangle = (u_1+v_1c_1^\dagger)(u_2+v_2c_2^\dagger)|0\rangle .
\label{eq:psi}
\end{equation}
The interference fringe appears in 
$P(x) \equiv \langle \psi|x\rangle \langle x|\psi\rangle$, 
the probability of finding
a particle at the location $x$ on the screen, as
\begin{eqnarray}
 P(x) &=& |u_2v_1\varphi_1(x)|^2 + |u_1v_2\varphi_2(x)|^2 \nonumber \\
      &+& 2|u_2v_1u_1v_2\varphi_1(x)\varphi_2(x)| \cos{\phi} \,, 
       \label{eq:P}
\end{eqnarray}
where $\varphi_j(x)\equiv \langle x|c_j^\dagger|0\rangle$ is the
wave amplitude of the single particle emitted from $S_j$. 
The phase shift $\phi$ of the interference fringe has two independent
contributions, 
 $\phi = \phi_B + \phi_0$,
where $\phi_B$ and $\phi_0$ originate from the interaction with the
localized magnetic field $\mathbf{B}$ and 
from the path difference, respectively. 
$\phi_0$ is independent of $\mathbf{B}$ and our main interest is $\phi_B$, 
which we analyze using both the potential-based and the LCFI approaches.

In the conventional potential-based theory, 
$\phi_B$ can be evaluated from
the interaction Lagrangian of Eq.~\eqref{eq:Lint0} (with $V=0$ in our case)
as
\begin{equation}
 \phi_B = \frac{1}{\hbar} \int_c L_{int} dt 
        = \frac{q}{\hbar c} \int_c \mathbf{A}\cdot d\mathbf{r} \,,
\label{eq:phiB-nl}
\end{equation}
where the integration is to be taken over the {\em open} path $c$ 
denoted in Fig.~1.
However, we face a fundamental problem here. 
$\phi_B$ in Eq.~\eqref{eq:phiB-nl}
is not well defined because the integration does not constitute a closed loop;
thus $\phi_B$ depends on the choice of gauge in $\mathbf{A}$.
That is, the nonlocal Lagrangian cannot predict the $\mathbf{B}$ dependence 
of the interference pattern derived in Eq.~\eqref{eq:P}. 


Next, let us evaluate $\phi_B$ by using the local interaction Lagrangian
of Eq.~\eqref{eq:Lintf}
in the presence of $\mathbf{B}$ (with $\mathbf{E}=0$). 
We find a gauge-independent result,
\begin{equation}
 \phi_B  = \frac{1}{\hbar} \int_c L_{int}^f dt 
         = \frac{1}{\hbar} \int_c \mathbf{\Pi}_q\cdot d\mathbf{r} \,.
\end{equation} 
The local theory predicts a well-defined $\phi_B$
as a function of the external magnetic field. 
For an ideal flux tube (of flux value $\Phi$) with
a negligible diameter, 
\begin{equation}
 \mathbf{\Pi}_q = \frac{q\Phi}{2\pi c\rho} \hat{\theta} \,,
\end{equation}
where $\rho$ and $\hat{\theta}$ are the distance from the flux tube
and the azimuthal unit vector of the position of $q$, respectively.
Hence, we obtain
\begin{equation}
 \phi_B = \frac{q\Phi}{2\pi\hbar c} \Delta\theta \,,
\label{eq:phi_B}
\end{equation}
where $\Delta\theta$ is the angle appearing in the open path $c$ (Fig.~1). 
An interesting point is that this phase is related to the
field angular momentum, $\mathbf{\cal L} = q\Phi/(2\pi c)$, 
and the phase shift can be rewritten as
$\phi_B = \mathbf{\cal L} \Delta\theta/\hbar$. 
$\phi_B$ is reduced to the AB phase 
for $\Delta\theta=2\pi$, as expected.

\section{Possible realistic experiment with Andreev interferometer}
An electronic interferometer 
is unsuitable for observing interference without a closed loop.
This is because the normal-state electron cannot be in a superposed state 
involving different numbers of particles,
and $u_jv_j=0$ ($j=1,2$) in Eq.~\eqref{eq:psi}. 
Therefore, no interference fringe
appears in the probability distribution, $P(x)$, 
of Eq.~\eqref{eq:P}. This problem
can be overcome by adopting the superconducting coherence in which gauge
symmetry breaking plays a major role~\cite{leggett91}.
Consider a schematic setup illustrated in Fig.~2. 
Each superconducting
source ($S_1$ and $S_2$) is tunnel-coupled to a normal electrode 
($N$). 
The separation of the two junctions
should be short enough to maintain the phase information of the condensates.
In fact, this is an Andreev interferometer~(see e.g.,
Refs.~\cite{petrashov93,petrashov95})
where the interference
arises from the two indistinguishable Andreev reflection~(AR)~\cite{andreev64}
processes.
A notable difference from the usual Andreev interferometer is that the
phase shift between the two superconducting condensate is controlled by
an external flux without forming a loop.
The ideal experimental procedure is as follows: 
(i) For a fixed magnetic
flux, an identical voltage $V$ 
is applied simultaneously to the two superconductors for a time interval
$\tau$, and the output current is measured at $N$; 
(ii) This process is
repeated many times, and
the average output current is recorded; 
(iii) Steps (i) and (ii) are repeated for different values of $\Phi$. 
The output current $I$ measured in this way 
is expected to show interference pattern as a function of $\Phi$,
as we describe below.

The Hamiltonian of this system is given by
\begin{equation}
 H = H_{S_1} + H_{S_2} + H_N + H_T ,
\end{equation}
where $H_{S_j}$ and $H_N$ represent the electrodes $S_j$ ($j=1,2$) and 
$N$, respectively. $H_T$ describes the tunneling process 
between each superconductor and the normal metal, 
\begin{equation}
 H_T = \sum_{j,k,p,\sigma} 
  \left( t_j^* c_{jk\sigma}^\dagger a_{p\sigma} 
       + t_j a_{p\sigma}^\dagger c_{jk\sigma}
  \right) \,,
\end{equation}  
where $c_{jk\sigma}$($c_{jk\sigma}^{\dagger}$) and 
$a_{p\sigma}$($a_{p\sigma}^\dagger$) annihilate (create) an electron
at $S_j$ and $N$, respectively ($k,p$, and $\sigma$ are momenta and
spin indices, respectively).

For $eV\ll \Delta$ ($\Delta$ being the superconducting gap parameter of 
$S_j$), quasiparticle transmission is prohibited and 
charge transport is mediated by the AR, 
which converts a Cooper pair from each $S_j$ to two electrons in 
$N$~\cite{andreev64}.
This corresponds to a transition from an initial state $|i\rangle$
to a final state $|f\rangle = a_{p\uparrow}^\dagger a_{-p\downarrow}^\dagger
|i\rangle$ via second-order processes in $H_T$, without 
real quasiparticle excitation in $S_j$.
Neglecting the voltage dependence (valid for $eV\ll\Delta$), 
this AR amplitude 
(from $S_j$ to $N$) is given by~\cite{schrieffer63}
\begin{equation}
 A_j = -\pi\rho_j |t_j|^2 e^{i\phi_j} \,, 
\end{equation}
where $\rho_j$ is the density of states of $S_j$.
Each AR is a coherent process 
conveying the phase information ($\phi_j$) of the superconducting condensate.
This phase coherence of the AR has been well established in real 
experiments with hybrid superconductor-normal metal structures 
(See e.g., Refs.~\cite{petrashov93,petrashov95}), and can be utilized for
our purpose. 

In our setup,
the two AR processes, each of which is
injected from $S_1$ or $S_2$, are 
indistinguishable, as the two processes share common initial and
final states: 
\begin{equation}
 |i\rangle = |\Psi\rangle_1\otimes|\Psi\rangle_2\otimes|FS\rangle_N \,,
\end{equation}
and $|f\rangle = a_{p\uparrow}^\dagger a_{-p\downarrow}^\dagger
|i\rangle$, respectively, where
$|\Psi\rangle_j$ ($j=1,2$) and $|FS\rangle_N$ denote the BCS condensate
of $S_j$ and Fermi sea of $N$, respectively.
Note that the same type of interference 
with independent sources
was observed for two independent Bose condensates~\cite{andrews97}, 
which also supports the feasibility of the proposed experiment.
During the time interval $\tau$ where the bias voltage $V$ is applied,
the output current exhibits interference as
\begin{equation}
 I = \frac{\pi e^2}{\hbar} V 
  \left[ \Gamma_1^2 + \Gamma_2^2 + 2\Gamma_1\Gamma_2\cos{(\phi_1-\phi_2)}
  \right],
\end{equation}
where $\Gamma_j = 2\pi \rho_j\rho_N |t_j|^2$ is the single-electron
hoping rate from $S_j$ to $N$, with $\rho_N$ denoting the density
of states of $N$.
The phase shift in the interference pattern of $I$ is given by
\begin{equation}
 \phi_1-\phi_2 = \phi_0 + \phi_B .
\end{equation} 
$\phi_0$ is a constant which is independent of the external 
$\mathbf{B}$, and
the field dependence of the phase shift is obtained in the local theory as
\begin{equation}
 \phi_B=\frac{e\Phi}{\pi\hbar c}\Delta\theta , 
\end{equation}
where the angle $\Delta\theta$ is determined by the geometry (as displayed
Fig.~1).
This phase shift is equivalent to that obtained in Eq.~\eqref{eq:phi_B}
with $q=2e$.

$\phi_B$ cannot be obtained from the potential-based theory 
as shown in Eq.~\eqref{eq:phiB-nl} 
because it does not predict a gauge-invariant result. 
Only the local approach predicts a well-defined $\phi_B$
for an interference produced by two indistinguishable AR processes.
Therefore, an interferometer of the type
shown in Fig.~2 can be adopted to conduct a realistic test of the quantum
locality in the electromagnetic interaction.
The different prediction results from the different viewpoint on the
nature of the interaction.
In LCFI theory, any quantum phase shift is a
result of the accumulation of the gauge-invariant local interactions, 
and the locality principle is preserved. 
In the potential-based approach, in contrast,
it is impossible to attribute the
phase shift to the interactions occurring at 
particular spacetime locations, and the interference without a closed loop
in the setup of Fig.~2 cannot be predicted in a consistent way.

\section{Discussion}
Notably, we can derive 
the general relation between the two interaction Lagrangians,
$L_{int}$(Eq.~\eqref{eq:Lint0}) and $L_{int}^f$(Eq.~\eqref{eq:Lintf}).
The potential-based interaction Lagrangian ($L_{int}$) 
for a moving charge can be rewritten as 
\begin{equation}   
 L_{int} = \frac{1}{c} \int \mathbf{j}_q\cdot \mathbf{A}\,d^3\mathbf{r}'
         - \int \rho_q V\,d^3\mathbf{r}' \,,
\label{eq:Lint}
\end{equation}
with the charge and current densities 
of a point charge given by $\rho_q = q\delta(\mathbf{r}'-\mathbf{r})$ and
$\mathbf{j}_q = q\dot{\mathbf{r}} \delta(\mathbf{r}'-\mathbf{r})$, 
respectively.
Applying the
Maxwell equations for both the charge $q$, 
\begin{equation} 
 \nabla\cdot \mathbf{E}_q = 4\pi\rho_q , \;\;\; 
 \nabla\times \mathbf{B}_q 
    - \frac{1}{c} \frac{\partial\mathbf{E}_q}{\partial t}  
  = \frac{4\pi}{c} \mathbf{j}_q \,,
\end{equation}
and the external fields,
\begin{equation}
 \mathbf{E} = -\nabla V - \frac{1}{c}\frac{\partial\mathbf{A}}{\partial t}, 
 \;\;\;
 \mathbf{B} = \nabla\times \mathbf{A} \,,
\end{equation}
we find
\begin{subequations}
\label{eq:L-Lf}
\begin{equation}
 L_{int}^f = L_{int} + \frac{dF}{dt} \,,
\label{eq:L-vs-Lf}
\end{equation}
where
\begin{equation}
 F = \frac{1}{4\pi c} \int \mathbf{E}_q\cdot\mathbf{A} \,d^3\mathbf{r}' \,.
\end{equation}
\end{subequations}

This is a remarkable result in that the standard potential-based picture 
can be transformed into 
the framework of the local interaction of the electromagnetic fields 
by discarding the total time derivative term, $dF/dt$. 
The transformed Lagrangian does not involve potentials, and therefore
both the quantum and
classical equations of motion of a charge can be described in terms of
the ``local" Lagrangian, $L^f\equiv L_0+L_{int}^f$, 
without relying on the electromagnetic potential.

Consequences of the relation derived in Eq.\eqref{eq:L-Lf} are as follows.
The dynamics of a system is, in general, fully determined by the action,
which is defined in each theory 
as $S = \int_{t_1}^{t_2} L \,dt$ and
$S^f = \int_{t_1}^{t_2} L^f \,dt$, respectively,
for a trajectory from the initial ($\mathbf{r}_1,t_1$) to
the final ($\mathbf{r}_2,t_2$) spacetime points.
Eq.~\eqref{eq:L-vs-Lf} gives
\begin{equation}
 S = S^f - F(\mathbf{r}_2,t_2) + F(\mathbf{r}_1,t_1) \,.
\label{eq:S_vs_Sf}
\end{equation}
With this relation, two actions yield
the same classical equation of motion. It is because
only the variation of the action, $\delta S$, is relevant in classical
Euler-Lagrange equation and we find $\delta S=\delta S^f$
from Eq.~\eqref{eq:S_vs_Sf}~(see e.g., Ref.~\cite{landau60}).
%

Quantum mechanical problems require more careful analysis, 
although the two approaches predict the same 
topological phase as shown in Eq.~\eqref{eq:phi-AB}.
In general,
any observable quantum effect is included in the
transition amplitude of a particle trajectory between the two arbitrary 
points in spacetime,
$(\mathbf{r}_1,t_1)$ and
$(\mathbf{r}_2,t_2)$, expressed as
\begin{equation}
 \langle \mathbf{r}_2,t_2|\mathbf{r}_1,t_1\rangle \propto
   \int_{\mathbf{r}_1}^{\mathbf{r}_2} {\cal D}[\mathbf{r}(t)]
   \exp{(iS[\mathbf{r}]/\hbar)} \,,
\label{eq:transition}
\end{equation}
which is obtained by summing over all possible paths (represented by
${\cal D}[\mathbf{r}(t)]$).
As one can find in Eqs.~\eqref{eq:S_vs_Sf} and \eqref{eq:transition}, 
the two interaction Lagrangians, 
$L_{int}$ and $L_{int}^f$, yield 
different phase factor for the transition amplitude unless 
$(\mathbf{r}_1,t_1)=(\mathbf{r}_2,t_2)$. 
This difference is manifested in the proposed interferometer without a loop.

We now address the last unresolved question:
the underlying reason why the two approaches, namely the LCFI and the
potential-based theories, 
provide different predictions for an interference without a loop.
To address this problem, we should note that
the potential-based Lagrangian of
Eq.~\eqref{eq:L} is derived under the condition that it satisfies 
the classical equation of motion with a proper consideration of the
symmetry. 
One may take an alternative procedure of derivation, based
on the interaction energy 
which corresponds to the work to establish the
system configuration~(see, e.g., Sec.~6.2 of Ref.~\cite{jackson62}).
We do not see any reason why this procedure is inappropriate in
quantum systems.

A different perspective 
on the Faraday's law of induction is helpful for our analysis,
which goes as follows. 
Variation of $\mathbf{B}$ 
induces an electric field (electromotive force (emf)). 
Interestingly, the induced electric field $\mathbf{E}$ can
be derived from the momentum conservation law. Consider a hypothetical
stationary charge $Q$ under $\mathbf{B}$. As far as the $\mathbf{B}$ is
confined in a finite region (as it should be), 
the total momentum of the system is
conserved under variation of $\mathbf{B}$. This can be expressed as
\begin{equation}
 Q\mathbf{E} + \dot{\mathbf{\Pi}}_Q = 0 \,, 
\label{eq:QE}
\end{equation} 
where
$\mathbf{\Pi}_Q$ is the field momentum generated by $Q$ and $\mathbf{B}$
(equivalent to $\Pi_q$ in Eq.~\eqref{eq:Pi} with $q=Q$).    
It is straightforward to show that the Faraday's law, 
$\nabla\times\mathbf{E} + \dot{\mathbf{B}}/c = 0$, 
can be derived from Eq.~\eqref{eq:QE}.

The magnetic interaction term is constructed as follows.
Suppose that only a moving charge $q$ with its velocity $\dot{\mathbf{r}}$ 
is initially present.
Activation and variation of $\mathbf{B}$ 
results in an induced electric field $\mathbf{E}$ satisfying Eq.~\eqref{eq:QE}.
To keep $\dot{\mathbf{r}}$ 
unchanged under the variation of $\mathbf{B}$, 
a work ($W_B$) against the emf is performed to the system at the rate
\begin{equation}
 \frac{dW_B}{dt} = -q \dot{\mathbf{r}} \cdot \mathbf{E}  \,.
\label{eq:dWdt}
\end{equation}
Combining Eqs.~\eqref{eq:QE} and \eqref{eq:dWdt}, we find
\begin{equation}
 W_B =  \dot{\mathbf{r}} \cdot \mathbf{\Pi}_q \,.
\label{eq:W}
\end{equation} 
This is the work required to establish $\mathbf{B}$ in the system, 
and constitutes the magnetic
interaction energy between the two entities (the first term of the right hand
side of Eq.~\eqref{eq:Lint2}). 
%
Similarly, we obtain the work ($W_E$) for establishing $\mathbf{E}$
as
\begin{equation}
 W_E = U_q \,,
\end{equation}
 where $U_q$ is the field interaction energy defined in
Eq.~\eqref{eq:U}.
Incorporating the Lorentz symmetry of the system, the two contributions
of the work, $W_B$ and $W_E$, constitutes the interaction Lagrangian
$L_{int}^f$ of Eq.~\eqref{eq:Lint2}.

The essential 
difference between the local and the nonlocal Lagrangians can be summarized
as follows.
In the local theory, the interaction Lagrangian is constructed
from the work necessary to establish the configuration. 
No gauge dependence is
included in this approach, and the interaction Lagrangian is given
without a potential (Eq.~\eqref{eq:Lintf}). 
In the conventional potential-based framework, in contrast, the interaction
Lagrangian of Eq.~\eqref{eq:Lint0} (or Eq.~\eqref{eq:Lint}) is 
derived under the condition that it satisfies the classical equation of
motion. 
It does not rely on the work performed to establish the configuration
and, thus, there is no way of assigning the local interaction energy. 
%
As we have shown above, observation of an interference without
a loop would confirm the locality of the interaction.

Finally, it should be noted that the previous experimental verification of 
the absence of the classical force in the AB setup~\cite{caprez07} is 
inadequate as regards our locality test. 
As the two different approaches yield the equivalent 
classical equation of motion (derived from Eq.~\eqref{eq:L-vs-Lf}),
the absence of force observed in 
Ref.~\cite{caprez07} can
be explained by both theories of the interaction. 
Therefore, the experimental
result in Ref.~\cite{caprez07} does not rule out any of the two approaches.
In fact, the classical lag predicted in Ref.~\cite{boyer73}, a
motivation of the experiment in Ref.~\cite{caprez07}, is found to be
erroneous which originates from the neglect of the Lorentz covariance
(See Ref.~\cite{kang13} for details).

\section{Conclusion} 
We have proposed a quantitative test of the locality
in the quantum electromagnetic interaction that induces the 
Aharonov-Bohm effect.
The equivalence and breakdown of the local and nonlocal theories  
have been analyzed.
The locality test can be performed by constructing 
an Andreev interferometer with the phase difference of the 
two superconducting condensates controlled by an external magnetic flux
without forming a loop.  
The two approaches provide different predictions for the interference fringe
and, therefore, 
a successful experiment is expected to resolve the issue of quantum locality 
in the electromagnetic interaction. 
Observation of the interference 
would discard the gauge-dependent potential-based theory, and therefore, 
rule out the dynamical nonlocality in quantum theory. 
This is of great significance
in our understanding of the nature of 
the electromagnetic interaction and the role of the potential. 

\section*{Acknowledgment}
This work was supported by the National Research Foundation of Korea (NRF)
(Grant No.~NRF-2015R1D1A1A01057325).
%
%

%
%
\section*{References}
\bibliographystyle{elsarticle-num}
\bibliography{references.bib}
%
\begin{figure}
 \includegraphics[width=3.2in]{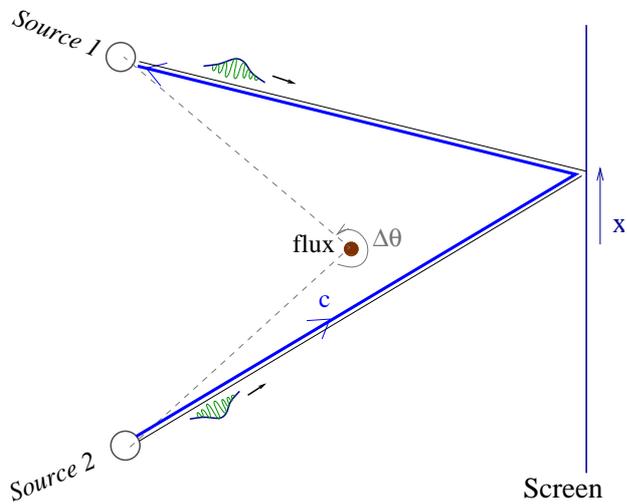}
\caption{Prototype setup for testing the quantum locality of 
the electromagnetic
interaction. The interference of a single charged particle ``coproduced"
by two independent sources with a localized magnetic flux can elucidate
the dynamical locality of the Aharonov-Bohm effect.
 }
\end{figure}
\begin{figure}
 \includegraphics[width=3.2in]{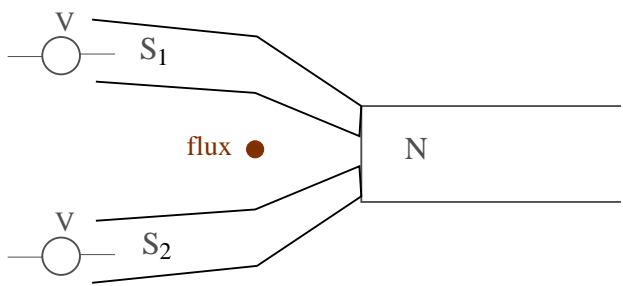}
\caption{A possible experimental test of  
the quantum locality of the electromagnetic interaction via an
Andreev interferometer.
Voltage $V$ is applied simultaneously to the two superconductors.
The Andreev reflection processes in the
two superconductor-normal metal contacts ($S_1$-$N$ and $S_2$-$N$)
are expected to show an interference
as a function of the external magnetic flux.
}
\end{figure}
\end{document}